\documentclass[a4paper]{jpconf}
\usepackage{graphicx}

\begin{document}

\title{Nuclear Reactions: A Challenge for Few- and Many-Body Theory}

\author{Ch. Elster, L. Hlophe}

\address{Department of Physics and Astronomy  and Institute of
Nuclear and Particle Physics, \\ Ohio University, Athens, OH 45701, USA}

\ead{elster@ohio.edu}

\begin{abstract}
A current interest in nuclear reactions, specifically with rare isotopes concentrates on their reaction with neutrons, in particular neutron capture. In order to facilitate reactions with neutrons one must use indirect methods using deuterons as beam or target of choice. 
For adding neutrons, the most common reaction is the (d,p) reaction, in which the deuteron breaks up and the neutron is captured by the nucleus. Those (d,p) reactions may be viewed as a three-body problem in a many-body context. This contribution reports on a feasibility study for describing phenomenological nucleon-nucleus optical potentials in momentum space in a separable form, so that they may be used for Faddeev calculations of (d,p) reactions.  
\end{abstract}

\section{Introduction}
Today the non-relativistic nuclear three-body system can numerically be solved 
with high precision in momentum~\cite{wgphysrep} as well as coordinate 
space~\cite{Barletta:2008hs}, including three-nucleon forces. Beyond this, 
three-body models can play an important role in the description of 
many-body problems. 

Since the full many-nucleon problem presently can only be solved for light 
nuclei~\cite{Pieper:2002ne,Pieper:2007ax,Navratil:2011ay,Navratil:2010jn}, a  
common approach to treating reactions involving heavier nuclei is concentrate on
fewer, relevant degrees of freedom and employ few-body methods. Such an approach 
is mostly justified for direct reactions, and may be expected to be reasonable 
if either the projectile fails to significantly penetrate the nucleus (i.e.
it is peripheral)  or if 
it penetrates, it moves very fast with respect to the nucleons bound in the nucleus. 
When a reaction is peripheral, it can be expected that only very few degrees of 
freedom of the nucleus may be excited due to the relatively low density at the nuclear surface. However, one can argue that in case of attractive optical potentials
a reaction will also be mostly peripheral. The projectile has a higher kinetic energy and a shorter wave length inside the nucleus, so that increased oscillations of the wave function inside the nucleus tend do suppress the contribution of the nuclear interior.

\begin{figure}[t]
\begin{center}
\includegraphics[width=108mm]{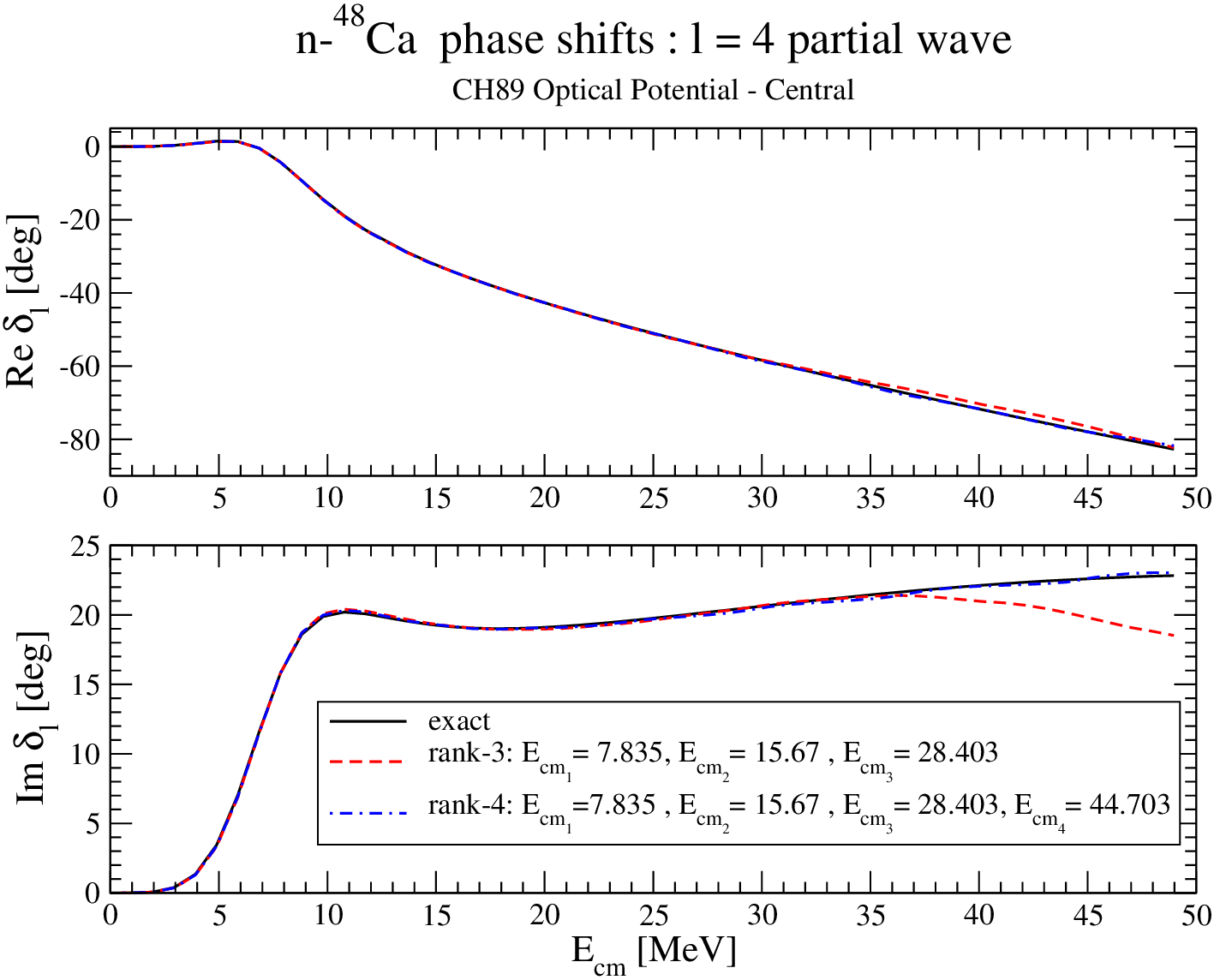}
\label{fig1}
\vspace{3mm}
\caption{The $l=4$ phase shift calculated from the central part of the CH89
optical potential~\cite{Varner:1991zz} for n+$^{48}$Ca (solid line) together
with a rank-3 (dashed line) and rank-4 (dash-dotted line) separable representation as
function of the c.m. energy.}
\vspace{5mm}
\includegraphics[width=106mm]{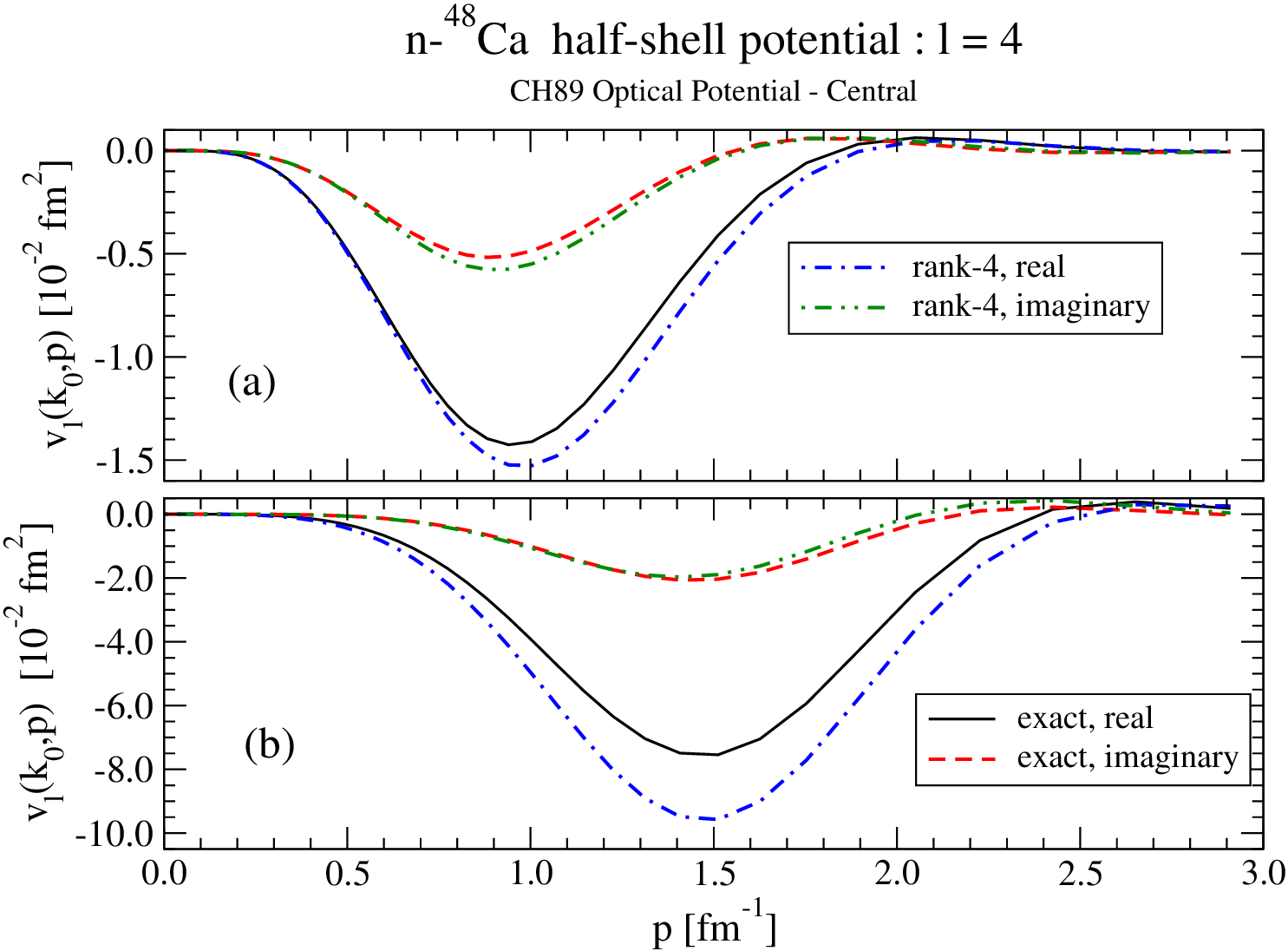}
\label{fig2}
\vspace{5mm}
\caption{The half-shell potentials $v_{l=4}(k_0,p)$ corresponding to the
CH89 central optical potential (exact) together with the rank-4 separable
representation at fixed momentum $k_0$ corresponding to $E_{c.m.}=4.897$~MeV
(a) and $E_{c.m.}=48.97$~MeV (b).}
\end{center}
\end{figure}

\section{(d,p) Reactions as Three-Body Problem}

Scattering and reaction processes induced by deuterons as projectile are
perhaps the most natural three-body problem in the realm of nuclear
reactions. The binding energy of the deuteron is so small that its
root-mean-square radius (rms) is significantly larger than the range of the
nuclear force, meaning that most of the time the neutron and the proton
inside the deuteron can be viewed as being outside the range of the
interaction. Thus, when a deuteron interacts with a compact nucleus, one may
expect that it will behave like a three-body system consisting of a proton
$p$, a neutron $n$, and a nucleus $A$. 

The most obvious three-body effects are rearrangement (stripping followed by
pickup) and breakup processes. In order to treat those processes as well as
elastic deuteron scattering on the same footing, 
deuteron-nucleus scattering should be treated as three-body problem making assuming that
the system can be described by a three-body Hamiltonian. It should be noted
that the extraction of an effective three-body Hamiltonian from the
many-body problem is non-trivial. A common approach is using as forces
between the neutron and proton (both interacting via the nucleon-nucleon
force) neutron and proton-nucleus optical potentials, which in turn are fitted
to a large body of elastic scattering data (e.g.~\cite{Weppner:2009qy,Varner:1991zz}). Since
these optical potentials are in general energy dependent, 
a common procedure is to take the nucleon-nucleus 
optical potentials at half the deuteron incident energy. With this
Hamiltonian Faddeev
type equations for (d,p) reactions can be set up and solved~\cite{Deltuva:2009fp}. 
Treating nuclear
reactions within a Faddeev three-body formulation is at the forefront of
developments in nuclear reactions. Existing
formulations need to be extended, since the original Faddeev description of
a three-body problem  does not consider possible internal
excitations of  the particles active in the scattering process.
In a (d,p) reaction with a nucleus, it is definitely possible that the
nucleus can be excited during the scattering process. When considering reactions involving
exotic nuclei, this may become even more urgent when the
exotic nuclei under consideration are loosely bound. 

A Faddeev formulation
of (d,p) reactions including excitations of the nuclear target has been
proposed in Refs.~\cite{Mukhamedzhanov:2012qv,Alt:2007wm}. Solving Faddeev
equations is most conveniently carried out in momentum space as coupled
integral equations, since those automatically contain the
boundary conditions. Thus all interactions in the three-body Hamiltonian
need to be given in momentum space. Specifically,
a momentum space Faddeev formulation requires as input from the sub-systems, 
which are here either the deuteron of the neutron-nucleus or proton-nucleus pair,
two-body transition
matrices (i.e. the interactions summed to all orders) for those sub-systems.
It is well known that the numerical effort in solving momentum space Faddeev integral equations
is reduced when separable representations of the sub-system transition amplitudes are used.
In the case of exited states of the target, separable interactions have the
additional important feature that for any resonance as well as for a 
bound state any two-body transition amplitude is separable. For this reason the 
formulation in Ref.~\cite{Mukhamedzhanov:2012qv} is based on using separable transition amplitudes for the sub-systems.

\begin{figure}[t]
\begin{center}
\label{fig3}
\includegraphics[width=115mm]{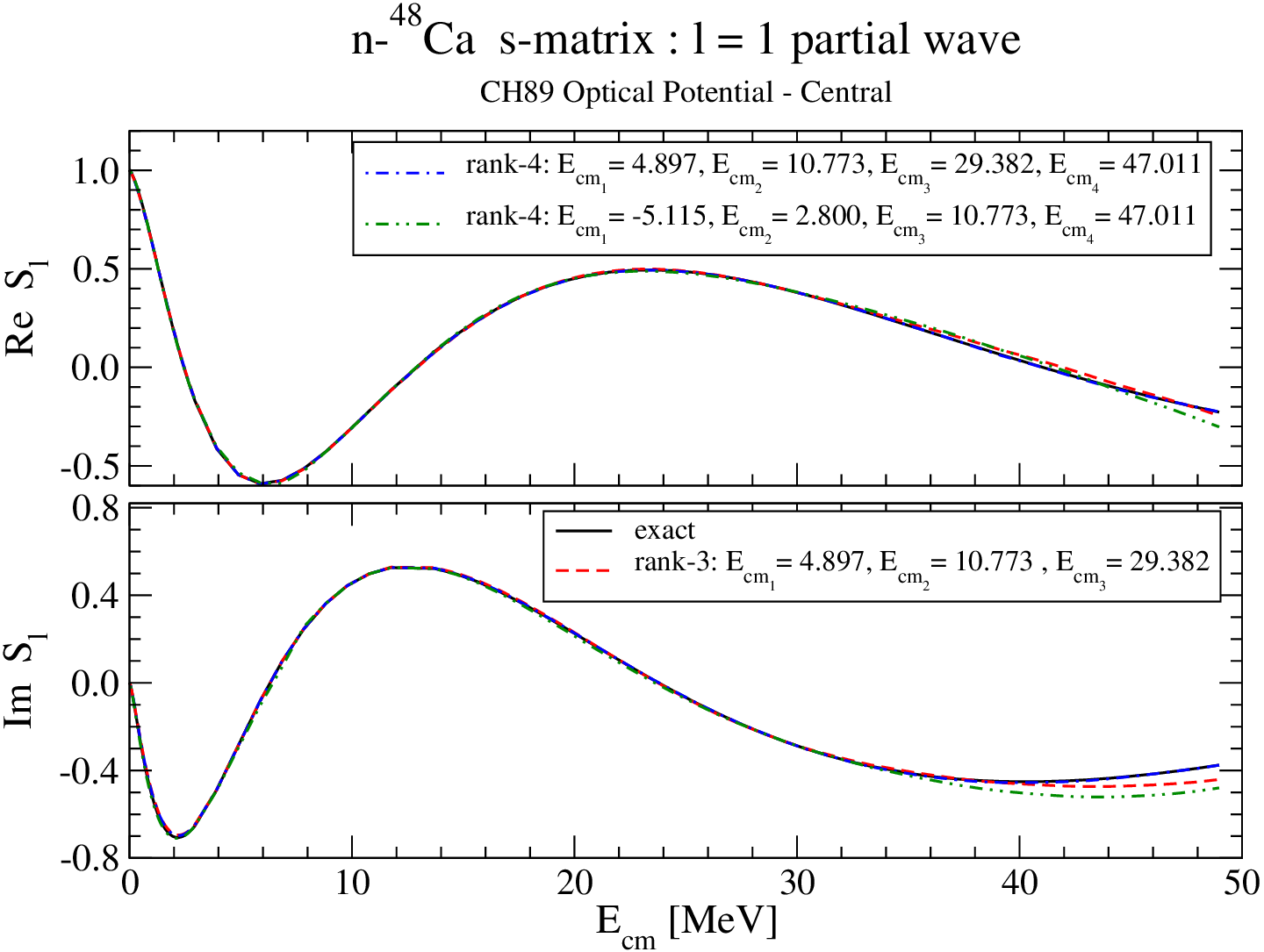}
\vspace{6mm}
\caption{The real (upper panel) and imaginary (lower panel) partial wave s-matrix
for $l=1$  calculated from the central part of the CH89
optical potential~\cite{Varner:1991zz} for n+$^{48}$Ca (solid line) as function 
the c.m. energy. A rank-3 (dashed) and two different rank-4 (dash-dotted and dash-double-dotted) are compared to the exact calculation.}
\end{center}
\end{figure}

\section{Separabilization of Nucleon-Nucleus Potentials}

Separable representations of nucleon-nucleon (NN) t-matrices are available since the 1980s (see e.g.
Refs.~\cite{Haidenbauer:1984dz,Haidenbauer:1986zza}). However,  as far as
neutron or proton-nucleus optical potentials are concerned, there only exist very 
few separable potentials~\cite{Alt:2007wm,MiyagawaK} for light nuclei,
which are based on Yamaguchi-type form factors. Those form factors may not
be well suited for
parameterizing optical potentials for heavier nuclei, for which excellent 
phenomenological descriptions in terms of Wood-Saxon functions exist (see e.g.
\cite{Varner:1991zz,Weppner:2009qy}). Considering the variety of nuclei for
which there is or will be experimental information available from (d,p)
reactions, one will need a separabilization procedure that is sufficiently
general so that it can be applied to a variety of nucleon-nucleus optical
potentials over a wide range of nuclei. In addition, transition amplitudes
calculated from phenomenological optical potentials need to be well
represented over a wider range of energies.

The separable representation of two-body interactions suggested by
Ernst-Shakin-Thaler~\cite{Ernst:1973zzb} (EST) looks well suited to achieve
this goal.  The basic idea of the EST separabilization of a two-body 
transition amplitude is that one
picks a fixed number of energies $E_{k_E}$ for which the transition amplitude is
exact, and all other energies are described by interpolation between those points
$E_{k_E}$ by the EST separable transition amplitude in which 
the half-shell amplitudes evaluated at $E_{k_E}$
serve as form-factors. The number of energy points $E_{k_E}$ determines the rank of the
separable t-matrix. The description of the original t-matrix can in principle be 
made successively more accurate with increasing support points $E_{k_E}$.

\begin{figure}[t]
\begin{center}
\includegraphics[width=115mm]{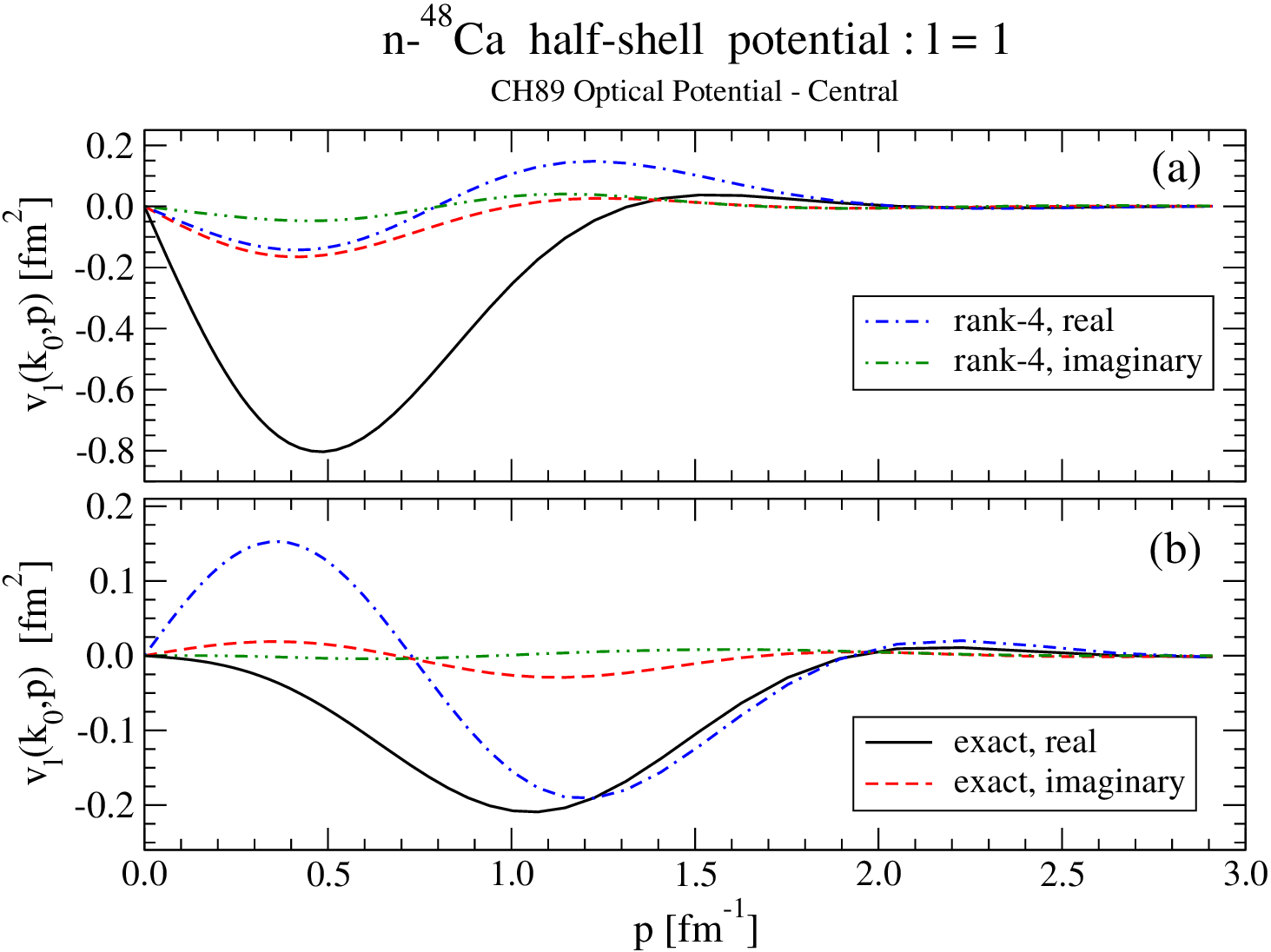}
\label{fig4}
\vspace{6mm}
\caption{The real (solid) and imaginary (dashed) partial wave projected 
half-shell CH89~\cite{Varner:1991zz} optical potentials for n+$^{48}$Ca scattering
 are shown for
fixed on-shell momentum $k_0$ as function of the half-shell momentum $p$. In (a) the on-shell momentum $k_0$ corresponds to $E_{c.m.}=4.897$~MeV, and in (b) to $E_{c.m.}=48.97$~MeV. A potential leading to a rank-4 separable representation of the corresponding s-matrix is shown by the dashed-dotted line (real part) and dashed-double-dotted line (imaginary part). 
}
\end{center}
\end{figure}

Though the EST scheme  was extensively used to represent NN 
t-matrices~\cite{Haidenbauer:1984dz,Haidenbauer:1986zza}, it has to our 
knowledge not yet been applied to  optical potentials for nucleon-nucleus 
scattering.  We report here on a study in which we
use the EST scheme to represent the partial wave projected 
 t-matrices for neutron
scattering from $^{48}$Ca based on the Chapel-Hill CH89 phenomenological
global optical potential~\cite{Varner:1991zz}. 
For this study we concentrate on the central part of the optical 
potential and neglect the spin-orbit terms. 

According to Ref.~\cite{Ernst:1973zzb} a separable potential of arbitrary rank is given as
\begin{equation}
{\bf V} = \sum_{i,j} v|\Psi_i\rangle \langle \Psi_i | M | \Psi_j\rangle \langle \Psi_j|v,
\label{eq:1}
\end{equation}
where $|\Psi_i\rangle$ stands for either a bound-state $|\Psi_B\rangle$ or a scattering state
$|\Phi^{(+)}_{k_E}\rangle$. If only bound states are considered, the standard unitary 
pole approximation (UPA) is recovered. The matrix $ M$ is defined and constrained by
\begin{equation}
\delta_{ik}=\sum_j \langle\Psi_i|M|\Psi_j\rangle \langle \Psi_j| v|\Psi_k\rangle =
 \sum_j \langle\Psi_i| v |\Psi_j\rangle \langle\Psi_j|M|\Psi_k\rangle.
\label{eq:2}
\end{equation} 
In order for the t-matrix elements  to have the same structure as the
Born term of the Lippmann-Schwinger integral equation, 
they must be of the form
\begin{equation}
{\hat t}(E) =  \sum_{i,j} v|\Psi_i\rangle \tau_{ij}(E) \langle \Psi_j|v \; .
\label{3}
\end{equation}
It can be shown~\cite{Ernst:1973zzb} that the quantities $\tau_{ij}(E)$ fulfill 
\begin{equation}
\sum_j \tau_{ij}(E) \; \langle \Psi_j|v -v g_0(E) v| \Psi_k\rangle = \delta_{ik} \;,
\label{eq:4}
\end{equation}
and thus can be calculated by inverting the matrix 
$\langle \Psi_j|v -v g_0(E) v| \Psi_k\rangle$ for given $|\Psi_{j,k}\rangle$,  so that
finally the momentum space t-matrix elements  are given as
\begin{equation}
{\hat t}(p,p',E) = \sum_{ij} \langle p| v_l|\psi_i\rangle \tau_{ij}(E) \langle \psi_j |v_l|p'\rangle
\label{eq:5}
\end{equation}

The central part of the CH89 optical potential for n+$^{48}$Ca scattering is attractive and supports either one ($l=2,3$) or two ($l=0,1$) bound states in the lower partial waves. Partial wave beyond $l=4$ do not support  bound states. For those partial waves the EST support points at with the exactly calculated half-shell t-matrix is used to construct the form factors for the separable expansion are all chosen to be at positive c.m. energies. In Fig.~\ref{fig1} the real and imaginary part of the $l=4$ phase shift are shown for a rank-3 and a rank-4 approximation and compared to the exactly calculated phase shift.  
The most prominent structure of the phase shift occurs roughly between 5 and 15~MeV c.m. energy. Thus two support points are needed to fit this structure and one additional at around 30~MeV to describe the flat behavior in that region. If one is interested to describe the phase shift at higher energies, then additional support points can be added without problem. 

In Fig.~2 the corresponding 
half-shell potentials $v_{l=4}(k_0,p)$, calculated using Eq.~(\ref{eq:1}),
 are shown for two different values of $k_0$ and compared to the original CH89 optical potential. 
For the upper panel (a) $k_0$ is fixed corresponding to $E_{c.m.}=4.897$~MeV and for the lower panel (b) to $E_{c.m.}=48.97$~MeV. Since the CH89 optical potential is energy dependent, so are the separable expansions. In the latter, the
form factors are energy dependent, which is formally allowed~\cite{Ernst:1974zzb}. The original, local Wood-Saxon potential  and the non-local rank-4 separable representation are amazingly similar half-shell. 

As example for a lower partial wave we consider $l=1$. This partial wave supports two bound states in its Wood-Saxon well, a ground state at $E_{B_1}=28.3$~MeV and an excited state at $E_{B_2}=5.1$~MeV. For a (d,p) reaction only the latter is relevant, since a neutron may be captured into this state. In Fig.~3 the partial wave $l=1$ s-matrix calculated from the CH89 cental optical potential is shown together with a selection of separable representations. If one is only interested in a representation of the s-matrix for scattering energies, then one can proceed in exactly the same fashion as in the higher partial waves that do not support bound states. A rank-3 and a rank-4 separable representation constructed in this fashion are shown in Fig.~3 as dashed and dash-dotted lines. The rank-4 expansion describes the s-matrix essentially perfectly up to 50~MeV. 
In a three-body calculation, which takes into account the capture of a neutron into the $l=1$ bound state close to threshold, the pole structure given by this bound state must be taken into account exactly. In a separable expansion this can be easily accomplished by using the bound state wave function as form factor (as done in UPA approximations) for one of the expansion terms. The dash-double-dotted line in Fig.~3 represents a separable expansion in which one of the terms contains the bound state wave function of the excited state as form factor. 
Obviously, the positive energy EST support points need to be readjusted. Their values are indicated in Fig.~3. We also tested if the description of the s-matrix depends on introducing the ground state wave function as an additional form factor. However, since its energy is quite far below threshold, the effect on the separable representation for positive energies is negligible.

In Fig.~4 the corresponding partial wave half-shell potentials $v_{l=1}(k_0,p)$ are shown for the central, local  CH89 optical potential as function of the half-shell momentum together a non-local, separable rank-4 potential obtained via  Eq.~(\ref{eq:1}).
As in Fig.~2, an on-shell point $k_0$ corresponding to $E_{c.m.}=4.897$~MeV is
chosen for panel (a) and to $E_{c.m.}=48.97$~MeV for panel (b). 
In this case the both half-shell potentials differ considerably from each other, despite describing the same s-matrix elements. 
The phenomenon that potentials may show considerable off-shell differences but still lead to very similar t-matrices (and thus s-matrices) has been studied for various NN potentials~\cite{Redish:1986jt}.  Since the information from two-body subsystems which enters Faddeev calculations only enters
via t-matrices, differences in potentials do not matter. If the fully off-shell t-matrices 
show substantial differences in momentum regions important to a Faddeev calculations will need to be investigated further. 

\section{Summary and Outlook}

For describing all aspects of (d,p) reactions on nuclei a three-body Faddeev formulation is currently the most desirable ansatz, since in a Faddeev formulation elastic scattering, 
breakup and rearrangement 
processes are treated on the same footing. Considerable  progress has been made in solving 
Faddeev equations in momentum space for (d,p) reactions~\cite{Deltuva:2009fp}, however there
are still major open issues. One of them is the inclusion of target excitations, i.e. the intrinsic excitation of one of the three-body constituents, which is not inherent in the original Faddeev formulation for three-body scattering. A formulation for (d,p) reactions including target excitations has recently  been proposed in Ref.~\cite{Mukhamedzhanov:2012qv}. which is based on
a separable representation of 
the transition amplitudes of the two-body sub-systems. While separable representations for the NN sub-system have been developed some time ago, for the neutron- and proton-nucleus sub-system corresponding representations do not exist. In a feasibility study based on neutron scattering off $^{48}$Ca and using only the central piece of a  Wood-Saxon based phenomenological 
optical potential fit~\cite{Varner:1991zz},
we show that for c.m. energies below 50~MeV the s-matrix can be well represented by rank-4 separable transition amplitudes. Future work will include a closer inspection of the fully off-shell 
behavior of the transition amplitudes derived from local optical potentials compared to their separable representations in momentum regions relevant for solving (d,p) Faddeev equations as well as
considering representations for the proton-nucleus optical potential.

\subsection*{Acknowledgments}
This work was performed in part under the
auspices of the U.~S.  Department of Energy under contract
No. DE-FG02-93ER40756 with Ohio University and
under contract No. DE-SC0004084 (TORUS Collaboration).

\end{document}